\documentclass[a4paper,fleqn,usenatbib,onecolumn]{mnras}
\usepackage[T1]{fontenc}
\usepackage{ae,aecompl}
\usepackage{graphicx}	
\usepackage{amsmath}	
\usepackage{amssymb}	
\title[Axisymmetric force-free magnetosphere] 
{Axisymmetric force-free magnetosphere 
in the exterior of a neutron star}
\author[Y. Kojima]{Yasufumi Kojima
\thanks{%
E-mail: ykojima-phys@hiroshima-u.ac.jp}\\ 
Department of Physics, Hiroshima University, Higashi-Hiroshima, Hiroshima 
739-8526, Japan}
\begin{document}
 \label{firstpage}
 \pagerange{\pageref{firstpage}--\pageref{lastpage}}
 \maketitle
%
\begin{abstract}
Magnetar magnetospheres gradually become twisted due to shearing 
by footpoint motion.
The axisymmetric solutions for the force-free field with a power-law 
current model are calculated by taking 
into account general relativistic effects.
Here we show how the magnetic energy and helicity are accumulated along 
a sequence of equilibria.
In a strongly twisted case, a magnetic flux rope, 
in which a large amount of toroidal field is confined, detaches 
in the vicinity of the star. 
It is found this kind of magnetic field line
structure is easily produced due to strong relativistic effects
compared with the results in flat spacetime.
Although the structure is not inherent to the relativity,
the confinement is possible even for a smaller power-law index.
There is an upper bound on the energy and helicity
stored in the magnetosphere. 
When the helicity is further accumulated, a catastrophic  
event such as a giant flare may occur.
\end{abstract}

\section{Introduction}
%
The magnetic fields of pulsars are inferred from the observed 
spin-down rate. The intensity at the surface exceeds the 
QED critical field $ 4.4 \times 10^{13}$G in a class of neutron stars  conventionally called magnetars
\citep{1995MNRAS.275..255T,1996ApJ...473..322T}, which typically exhibit energetic behaviors. 
The observed X-ray luminosity is much higher than the spin-down luminosity 
in both anomalous X-ray pulsars (AXPs) and soft $\gamma$-ray repeaters (SGRs).
They are very variable, and sometimes exhibit bursts, unlike normal pulsars.
In giant flares of SGRs, huge amounts of energy $10^{44}$-$10^{46}$ erg are 
released in an instant. 
The activity of magnetars is widely believed to be supplied 
by their strong magnetic field of $10^{14}$-$10^{15}$G.
Recent observations have hindered a definite neutron star classification.
There are exceptional bursting sources with normal magnetic field
($ < 4\times 10^{13} $G), i.e, `low-field' magnetar \citep{2010Sci...330..944R}.
Outbursts are observed in a rotation-powered pulsar\citep{2016ApJ...829L..25G}. 
See, for example, \cite{2008A&ARv..15..225M,2015RPPh...78k6901T}
for a review of magnetars.

 Two possibilities are proposed for the energy storage 
prior to magnetar outbursts to explain the relevant phenomena:
storage in the magnetar crust or in the magnetosphere \citep{2002ApJ...574..332T}.
The latter model is discussed in terms of similarity with solar flares
\citep{2003MNRAS.346..540L,2006MNRAS.367.1594L,2007ApJ...657..967B}.
In the solar flare model
\citep[e.g.,][]{1984ApJ...283..349A,1994ApJ...430..898M},
the energy is quasi-statically stored by thermal motion 
at the surface, and is suddenly released as large-scale eruptive coronal 
mass ejections.
The energy is dissipated via a magnetic reconnection associated with 
the field reconfiguration.
Analogous energy buildup and release processes 
may be relevant to the magnetar giant flares, although the energy scale 
differs by many orders.

The twisted magnetosphere formed by current flow is studied in  
the calculations of synthetic spectra 
\citep{2006MNRAS.368..690L,2007ApJ...660..615F,2008MNRAS.386.1527N}
and polarization \citep{2014MNRAS.438.1686T}.
In their calculations, the external global magnetosphere
is described by a self-similar bipolar field.
The solution of the force-free field is given by 
\citet{1994MNRAS.267..146L,2002ApJ...574..332T}, and  
is extended to a self-similar one with
higher order multipoles \citep{2009MNRAS.395..753P}.
Recently, \citet{2016MNRAS.462.1894A}
numerically constructed the force-free magnetosphere of a magnetar, 
which is smoothly matched to a vacuum solution at large distance.  
The magnetar magnetospheres are also studied
to be matched with the interior magnetic field
\citep[e.g.,][]{2011A&A...533A.125V,
2014MNRAS.437....2G,2014MNRAS.445.2777F,2015MNRAS.447.2821P}.
The exterior magnetic fields depend
on the interior ones through the boundary condition.
The magnetic field structure 
evolves on a secular timescale 
by Hall drift, ambipolar diffusion, or some other mechanism
\citep[e.g.,][]{1992ApJ...395..250G,2004MNRAS.347.1273H,2012MNRAS.421.2722K,
2013MNRAS.434..123V,2014MNRAS.438.1618G,2015PhRvL.114s1101W}.
A quasi-steady shearing motion  at the footpoint of
a magnetic field twists the exterior field, and the stored energy 
gradually increases.
When the state exceeds a threshold, the energy is abruptly released 
on a much shorter dynamical timescale, leading to energetic flares.
One important fundamental process
is the accumulation of magnetic helicity in the corona before 
the outbursts.
The magnetic energy also builds up as a natural product of the evolution of
the magnetic helicity.
The role of helicity accumulation in the solar corona
has been discussed \citep{2006ApJ...644..575Z}.  
The same process should be considered in the framework of 
general relativity, since
the buildup of magnetic energy occurs very near the neutron star surface.

\citet{2015MNRAS.447.2821P}
obtained axially symmetric equilibrium by solving
relativistic MHD equations both in the interior and 
in the exterior of a neutron star.
The magnetosphere is modeled by filling low-density plasma,
and the magnetic field is twisted by the current.
Magnetic pressure is much larger than hydrodynamical pressure,
and so the force-free approximation may be applicable.
We here provide a systematic study of a twisted force-free 
magnetosphere by limiting our consideration to the exteriors.
The force-free magnetosphere around a black hole
has been examined \citep[e.g.,][]{2004ApJ...603..652U}.
Our problem is simpler because the
charge density associated with plasma rotation is ignored.
The magnetars are slow rotators, and the current flow is
more important.

  In this paper, we examine the effect of general relativity
on the force-free magnetosphere in the vicinity of 
neutron stars. We consider static and axially symmetric 
magnetic fields in the exterior of a non-rotating compact star.
The spacetime is described by the Schwarzschild metric, 
and a family of equilibrium solutions are calculated 
assuming that the current function is given by a 
simple power-law model, for which solutions in a flat spacetime
have been well studied by \citet{2004ApJ...606.1210F,2006ApJ...644..575Z,
2012ApJ...755...78Z} in the context of solar flares.
This will clarify the effect of general relativity.
Equations relevant to non-rotating force-free  magnetosphere
in a Schwarzschild spacetime 
are discussed briefly in Section 2.
We numerically solve the so-called Grad--Shafranov equation\footnote{
Although this name is used in many papers, 
a work by L$\ddot{\rm u }$st and Sch$\ddot{\rm u }$ter
prior to Grad and Shafranov, 
was not recognized, as noted by \citet{2016MNRAS.462.1894A}. }
with a power-law model.
In Section 3, the results are given 
for comparison with those obtained in flat spacetime.
Finally, our conclusions are given in Section 4.
We use  geometric units of $c=G_{\rm N}=1$.

\section{Equations}
  \subsection{Magnetic fields}
In this section, we briefly summarize the Maxwell equations in curved 
spacetime outside of a non-rotating compact object.
The metric in the exterior of a mass $M$ in spherical coordinates 
is given by
\begin{equation}
ds^2=-\alpha^2dt^2+\alpha^{-2}dr^2
       +r^2d\theta^2 +\varpi^2 d\phi^2,
\end{equation}
where
\begin{equation}
\alpha^2=1-\frac{2M}{r},
~~
\varpi =r\sin \theta .
\end{equation}
We consider the static magnetic configuration. 
Using vector analysis in 3-dimensional curved space,
the relevant equations are expressed as
\citep[e.g.,][]{1986bhmp.book.....T}
\begin{equation}
\vec{\nabla}\cdot\vec{B}=0,
\label{eqnMxw2}
\end{equation}
\begin{equation}
\vec{\nabla}\times(\alpha{\vec{B}})= 4\pi\alpha\vec{j} .
\label{eqnMxw4}
\end{equation}
The magnetic field for the axially symmetric case is given by
two scalar functions $G(r, \theta)$ and $S(r, \theta)$:
\begin{equation}
\vec{B}=\vec{\nabla}\times \left(
\frac{G}{\varpi}\vec{e}_{\hat{\phi}}
\right)
+\frac{S}{\alpha \varpi}\vec{e}_{\hat{\phi}}
=
\frac{\vec\nabla{G}\times\vec{e}_{\hat{\phi}}}{\varpi}
+\frac{S}{\alpha \varpi}\vec{e}_{\hat{\phi}},
\label{eqnDefBB}
\end{equation}
where $\vec{e}_{\hat{\phi}} \equiv \varpi^{-1} \partial _{\phi}$
is a unit vector in the azimuthal direction.
The magnetic field (\ref{eqnDefBB}) automatically satisfies the 
constraint equation (\ref{eqnMxw2}), and the components are 
explicitly written as
\begin{equation}
[B_{\hat{r}},B_{\hat{\theta}}, B_{\hat{\phi}}]
=\left[\frac{G,_\theta}{r \varpi},
~
 -\frac{\alpha G,_r}{\varpi},
~
 \frac{S}{\alpha \varpi}\right].
\label{Bcomp}
\end{equation}
We introduce a vector potential ${\vec A}$
derived by functions $F(r, \theta)$ and $G(r, \theta)$
to express the magnetic field as
${\vec B} ={\vec \nabla} \times {\vec A}$:
\begin{equation}
\vec{A}=\vec{\nabla}\times \left(
\frac{F}{\varpi}\vec{e}_{\hat{\phi}}
\right)
+\frac{G}{\varpi}\vec{e}_{\hat{\phi}} 
=
\frac{{\vec \nabla} F\times\vec{e}_{\hat{\phi}}}{\varpi}
+\frac{G}{\varpi}\vec{e}_{\hat{\phi}} .
\label{eqnDefAA}
\end{equation}
There is gauge freedom in the potential ${\vec A}$.
The electromagnetic fields are unchanged by
${\vec A} \to {\vec A}^\prime= {\vec A}+{\vec \nabla} \Lambda$.
In the axially symmetric case ($\partial_\phi=0$), $G(=A_{\phi})$ is
gauge-invariant, and has a special meaning of magnetic flux.
The poloidal part of the vector potential is determined by
\begin{equation}
\alpha \frac{\partial}{\partial r}
\left( \alpha\frac{\partial F}{\partial r} \right)
+\frac{ \sin\theta}{r^2} \frac{\partial}{\partial \theta}
\left( \frac{1}{\sin \theta}\frac{\partial F}{\partial \theta} \right)
= -\frac{1}{\alpha}S.
\label{eqn:BFF}
\end{equation}
The force-free magnetic field
${\vec j}\times {\vec B}=0$ means
that current flows along the magnetic field lines.
The poloidal component of eq.(\ref{eqnMxw4})
(the Biot-Savart law) 
is reduced to $\nabla S  || \nabla G$, 
that is, $S=S(G)$, and $ 4\pi \alpha {\vec j}_p =S^\prime {\vec B}_p$,
where $S^\prime=dS/dG$.
This means that the poloidal current flows along a 
line of constant $S$. 
A factor $\alpha$ is needed in front of the current,
since not $ {\vec j}_p $ but $ \alpha {\vec j}_p $ 
should be proportional to ${\vec B}_p$ from 
the charge conservation
$\partial \rho_{e}/\partial t + {\vec \nabla} \cdot  (\alpha {\vec j})=0$.
Using this form, the azimuthal component is given by
$ 4\pi \alpha  j_\phi=S^\prime  B_\phi$
$=SS^\prime/(\alpha \varpi)$.
The azimuthal component of eq. (\ref{eqnMxw4})
therefore reduces to the so-called Grad-Shafranov equation:
\begin{equation}
\alpha^2 \frac{\partial}{\partial r}
\left( \alpha^2\frac{\partial G}{\partial r} \right)
+\frac{ \alpha^2 \sin\theta}{r^2} \frac{\partial}{\partial \theta}
\left(\frac{1}{\sin \theta}\frac{\partial G}{\partial \theta}\right)
= -SS^\prime .
\label{eqn:FF}
\end{equation}
This equation can be also derived 
in the non-rotating limit of the equation
used for black hole magnetospheres 
\citep{2004ApJ...603..652U}.

Magnetic energy is given by
integrating over 3-dimensional volume 
\begin{equation}
E_{\rm EM} = \int \frac{\alpha B^2}{8\pi} \sqrt{g_3}d^3x
= \frac{1}{4}\int \left[ \left( \alpha  
\frac{\partial G}{\partial r} \right)^2
+ \left( \frac{1}{r}\frac{\partial G}{\partial \theta} \right)^2
+\left(\frac{S}{\alpha} 
\right)^2  \right] 
\frac{dr d\theta }{\sin \theta},
\label{eqn:Eng}
\end{equation}
where $ \sqrt{g_{3}} (= \alpha^{-1} r^2 \sin \theta)$ is a 
determinant of 3-dimensional space metrics.
We consider the energy deposited outside the star, so that
the integration range is limited to $r\ge R$.
In eq. (\ref{eqn:Eng}), a factor $\alpha$ in front of
$B^2$ may be understood by considering the Maxwell equations
in curved spacetime. Equivalently, 
the expression (\ref{eqn:Eng})
can also be obtained by $\int T^{t} _{t}  \sqrt{-g_{4}} d^3x$
in terms of energy momentum tensor $T^{t} _{t} $ and 
determinant of 4-dimensional spacetime metrics $\sqrt{-g_{4}}$.
Helicity is obtained by integrating the product of
two vectors, namely, ${\vec A}$ and $ {\vec B} $.
Total relative helicity $H_{\rm R} $ defined by 
the integration range $r\ge R$ is 
\begin{equation}
H_{\rm R} 
=
  \int {\vec A}\cdot {\vec B} \sqrt{g_3}d^3x
= 4\pi\int  \frac{ GS}{\alpha^{2}}
\frac{dr d\theta }{\sin \theta} ,
\label{eqnHR}
\end{equation}
where the poloidal part derived by $F$ in the vector potential is 
eliminated by integrating by parts with eq.(\ref{eqn:BFF}).
In the limit of flat spacetime ($\alpha \to 1$), 
eq. (\ref{eqnHR}) reduces to 
the equation given by \citet{2006ApJ...644..575Z}.
When $S$ is a linear function of $G$,
eq. (\ref{eqn:FF}) is derived by minimizing the 
magnetic energy (\ref{eqn:Eng}) for fixed helicity (\ref{eqnHR}).
Thus force-free equilibrium with constant $S^\prime $
is a minimum energy state.
This is a relativistic extension of the work by
\citet{1958PNAS...44..285C}.
In presence of the toroidal component,
the magnetic field lines deviate in the azimuthal direction. 
We introduce a twist angle, which is closely related to helicity.
The magnetic field lines satisfy the relationship:
\begin{equation} 
\frac{\alpha^{-1}dr}{B_{\hat{r}}}=\frac{rd\theta}{B_{\hat{\theta}}}=
\frac{\varpi d\phi}{B_{\hat{\phi}}}.
\end{equation}
The twist angle along the line
from $(r_1,\theta_1) $ to $(r_2,\theta_2) $ 
is evaluated by one of the following expressions:
\begin{equation}
\Delta \phi = \int _{r_1} ^{r_2} 
\frac{r S}{\alpha^2 \varpi G_{,\theta}} dr
= -\int _{\theta_1} ^{\theta_2} 
\frac{r S}{\alpha^2 \varpi G_{,r}} d \theta .
\label{eqn:wistang}
\end{equation}

 \subsection{Model and numerical methods}

The potential dipole field, that is, the solution of eq. (\ref{eqn:FF}) 
for $S=0$ with magnetic moment $\mu$, 
is given elsewhere \citep[e.g.,][]{1974PhRvD..10.3166P,
1983ApJ...265.1036W,1999A&A...352..211K}:
\begin{eqnarray}
 \label{eqnpuredip}
G_{\rm d} &=& - \frac{3 \mu r^2}{8 M^3}\left[
\ln\left(1-\frac{2M}{r}\right) 
+\frac{2M}{r}
+\frac{2M^2}{r^2}
\right]\sin^2 \theta
\\
&\approx &
\frac{ \mu}{r}\left[
1+\frac{3M}{2r}+\frac{12M^2}{5r^2} +\cdots
\right]\sin^2 \theta,
\nonumber
\end{eqnarray}
where the first expression is the exact solution
and the second is an approximation in the weak gravity regime
$M/r \ll 1$.
Using magnetic dipole moment $\mu$,
the surface field strength is given by 
$ B_0 \equiv \mu R^{-3}$, where $R$ is the stellar radius.
The magnetic field at the stellar surface
is $B_{\hat{r}} \approx (1 +3M/(2R)) \times 2B_0 \cos \theta$
and $B_{\hat{\theta}} \approx (1 +2M/R) \times B_0 \sin \theta$
from eq. (\ref{Bcomp}).
The field strength at the pole is $2B_0$ in flat spacetime, but
increases by some factor in a relativistic treatment
for a fixed $\mu$.
The correction depends on the direction, but  
is typically a factor $\approx 1.5$ for $M/R \approx 0.25$.
The magnetic energy $E_{0}$ stored in the exterior $(r \ge R)$ 
for the solution (\ref{eqnpuredip}) is expressed by 
\begin{equation}
E_{0}  = \frac{  B_0 ^2 R^3 }{3 }\left[
1 + \frac{5 M}{2 R}+ \frac{51 M^2}{10 R^2}
 +\cdots \right].
\end{equation}
The energy increases by a factor  $\approx 2$
for $M/R \approx 0.25$ and a fixed $\mu$.
In this paper, we use $B_0$ and $R$ for normalization, because 
$B_0$ is usually estimated from 
measurement of spin period and its time derivative of pulsars.

Our concern is how energy and helicity change by
twisting the magnetic field in the exterior of a relativistic star.
We adopt a power-law model for $S$ as
\begin{equation}
S=\left(\frac{2\gamma}{n+1} 
\right)^{1/2}G^{(n+1)/2} .
\label{powerlaw}
\end{equation}
The right-hand side of eq. (\ref{eqn:FF}) thus becomes 
$-{\mathcal S} \equiv -SS^{\prime} =-\gamma G^{n}$,
where $\gamma$ is constant, and $n$ 
is assumed to be an odd integer 
in order to ensure that $S^2$ is positive definite
independent of the sign of $G$. 
The force-free magnetosphere with the power-law
model has been extensively studied in flat spacetime for the solar 
flare model \citep{2004ApJ...606.1210F,2006ApJ...644..575Z,
2012ApJ...755...78Z}.
There is no solution different from 
that of a pure dipole in the case of index $n< 5$ in the flat 
spacetime treatment.
We calculate the models with $n=5$ and 7 in this paper.
A continuous sequence of solutions is thus generated by 
varying a free parameter.
However, the sequence may not necessarily correspond to evolutionary track.
Physical evolution of the magnetosphere is not obtained until we know 
the evolution of footpoint motion.
As a possible scenario, suppose that the magnetosphere gradually changes 
by current flow from the surface.
The magnetic field is twisted, and the field topology changes by helicity 
accumulation.
Not only the parameter $\gamma$ but also the functional form $\gamma G^n$ 
should be changed in the power-law description in order to 
describe the evolution.
In spite of this fact, the sequence of solutions with a fixed index $n$
is easily constructed with an increase of magnetic twist,
and provides an understanding of the structure change by helicity accumulation.
The boundary conditions for solving eq. (\ref{eqn:FF}) are as 
follows.
At the stellar surface $r=R$, 
we assume that the magnetic function $G$ is given by 
that of a dipole in vacuum,
that is, eq. (\ref{eqnpuredip}): 
\begin{equation}
G = G_{\rm d}(R, \theta) .
\label{eqn:bcsurf}
\end{equation}
At asymptotic infinity ($r \to \infty $),
the function is expected to decrease as $G \propto r^{-1} $.
At the polar axis, the function is expected to satisfy the regularity condition,
that is, $ G=0$ at $\theta =0$ and $ \pi$.

Equation (\ref{eqn:FF}) with the power-law model (\ref{powerlaw}) is
non-linear except for $n=1$, and an iterative method is therefore
necessary for solving it.
An initial guess for the function $G_{(1)}$ is assumed.
Using this, the right hand side of eq. (\ref{eqn:FF})
is calculated, and a new function
$G_{(2)}$ is solved from the source term with appropriate boundary 
conditions. The procedure is repeated until
convergence, say, 
$||G_{(k+1)}-G_{(k)}||$ $<\varepsilon $, 
where $\varepsilon $ is a small number, and 
the left hand side is evaluated by discrete points
$(r_i,\theta _j)$ inside the domain: 
$||G_{(k+1)}-G_{(k)}|| \equiv$ 
$[ \sum_{i,j} (G_{(k+1)}(r_i,\theta _j)-G_{(k)}(r_i,\theta _j))^2 ]^{1/2}$
$ / [ \sum_{i,j} G_{(k)}(r_i,\theta _j)^2 ]^{1/2}$.
The iteration scheme may not necessarily lead to a convergent
solution, but we experimentally 
have a convergent solution for certain initial trials.

One of the numerical methods for solving eq. (\ref{eqn:FF})
is discretizing the equation \citep[e.g.,][]{2016MNRAS.462.1894A}.
Here we use another approach by polynomial 
expansion \citep[e.g.,][]{2014MNRAS.445.2777F}.
The magnetic function $G$ is expanded by Legendre polynomials 
$P_{l}(\theta)$:
\begin{equation}
G(r,\theta )=-\sum_{l \ge 1}g_{l}(r)\sin \theta 
\frac{d P_{l}(\theta )}{d\theta },
   \label{Sexpd.eqn}
\end{equation}
where radial functions $g_{l}$ 
in the case of the homogeneous equation of eq. (\ref{eqn:FF}) satisfy:
\begin{equation}
\frac{d}{d r}
\left( \alpha^2\frac{d}{dr}g_{l} \right)
-\frac{ l(l+1) }{r^2} g_{l}= 0.
\label{eqn:radial}
\end{equation}
Two independent solutions are obtained as $r^{l+1}$ and $r^{-l}$ 
in flat spacetime.
The solutions can be expressed by hypergeometric function in a
Schwarzschild spacetime. 
They are constructed by numerically solving eq. (\ref{eqn:radial}).
We denote two independent solutions by $ g^{-} _{l}$ and $ g^{+} _{l}$.
They are characterized by asymptotic behavior at large $r$:
$ g^{-} _{l} \to r^{-l}$ and $g^{+} _{l} \to r^{l+1}$.
The general solution of eq. (\ref{eqn:FF}) can be written 
using constants $a_l, b_l$ and a particular solution $q_{l}(r)$:
\begin{equation}
G(r,\theta )=-\sum_{l \ge 1}( a_l g^{-} _{l}(r) + b_l g^{+} _{l}(r)+ q_{l}(r) )
\sin \theta 
\frac{d P_{l}(\theta)}{d\theta }.
\end{equation}
When the source term
$-{\mathcal S} = -SS^{\prime} $ in eq. (\ref{eqn:FF})
is assumed to be given by an iterative process, 
the inhomogeneous solution $q_{l}$ 
can be solved by the Green function method:
\begin{equation}
q_{l}(r) = g^{-} _{l} (r) \int_{R} ^{r} 
 \frac{g^{+} _{l} (r^{\prime})S_{l}(r^{\prime})}{W_l(r^{\prime})} dr^{\prime} 
+ g^{+} _{l} (r) \int_{r} ^{r_\infty} 
\frac{g^{-} _{l}(r^{\prime})S_{l}(r^{\prime})}{W_l(r^{\prime})}dr^{\prime},
 \label{eqn.specl}
\end{equation}
where $W_l$ is  a radial function defined by
\begin{equation}
W_l \equiv 
\alpha^{4}\left[ g^{-} _{l}\frac{dg^{+} _{l}}{dr}
 -g^{+} _{l}\frac{dg^{-} _{l}}{dr}
\right]
\end{equation}
and $S_{l}$ is the $l$-th component of the source term ${\mathcal S} $:
\begin{equation}
S_{l}(r) = 
\frac{2l+1}{2l(l+1)}\left[ \int_{0} ^{\pi}  {\mathcal S}(r,\theta)
\frac{d P_{l} (\theta)}{d\theta} d\theta \right] .
\end{equation}
Although the upper limit $r_\infty$ of the integral should be $\infty$,
$r_\infty /R >$ 20-30 is sufficient in the actual 
numerical calculation.
The inhomogeneous solution in eq. (\ref{eqn.specl})
satisfies $q _{l} \propto r^{-l}$ in the limit of $r \to \infty$, 
so that all coefficients $b_{l}$ 
in front of growing functions should be zero.
The condition at the surface, eq. (\ref{eqn:bcsurf})
determines the coefficients $a _{l}$.
In the numerical calculation, we truncate the polynomial expansion 
in eq. (\ref{Sexpd.eqn}) up to $l_{\rm max}=$64-128.
The numerical results are almost unchanged by further increasing it.

\section{Numerical results}
\begin{table}
\caption{ Model and result.
A family of the solutions is
characterized by power-law index $n$
and relativistic factor $M/R$.
The magnetic energy for the potential
field is $E_0$, and maximum energy increment $\Delta E$  
from $E_0$, relative helicity $H_{\rm R}$
and twist angle $\Delta \phi _{\rm max}$ are also listed. 
}
\begin{center}
\begin{tabular}{ccccccc}
\hline\hline
Model &  $n$ &$M/R$ &
 $E_0/(B_0 ^2 R^3)$ & $(\Delta E/E_0) _{\rm max}$  
& $(H_{\rm R}/(4\pi B_0 ^2 R^4)) _{\rm max}$  
& $\Delta \phi_{\rm max}$  
 \\ \hline
F5 & 5 & 0    & 0.33 & 0.34  & 0.74  & 1.05  \\ 
F7 & 7 & 0    & 0.33 & 0.58  & 0.77  & 1.49  \\ 
G5 & 5 & 0.25 & 0.71 & 1.32  & 3.25   & 1.38 \\ 
G7 & 7 & 0.25 & 0.71 & 2.78  & 3.50   & 0.88 \\ 
\hline\hline
\end{tabular}
\end{center}
\label{Table:I}
\end{table}

\begin{figure}
\centering
\includegraphics[scale=0.45]{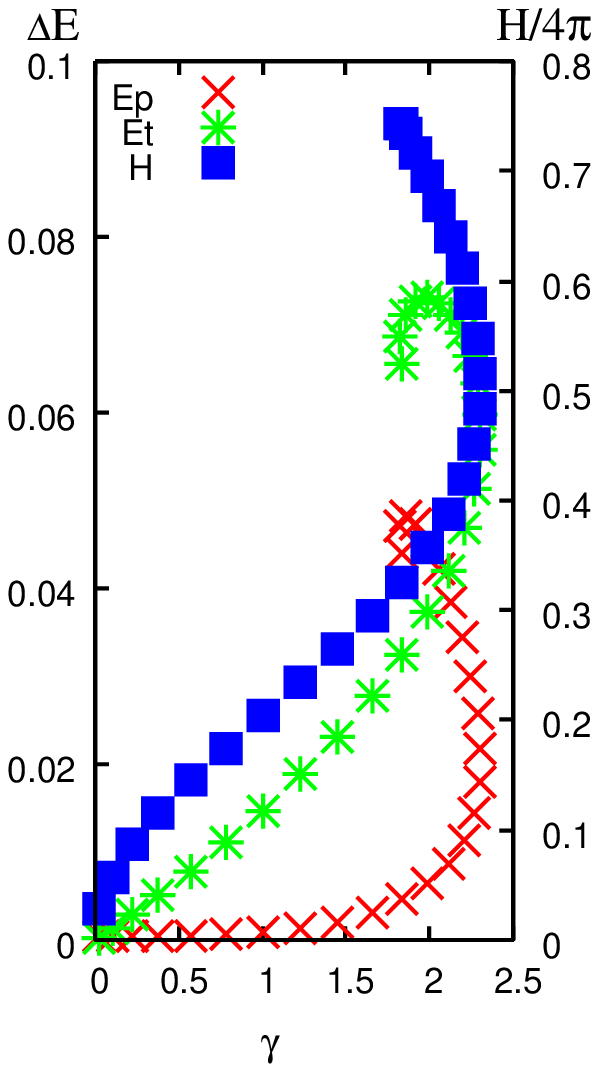}
\includegraphics[scale=0.45]{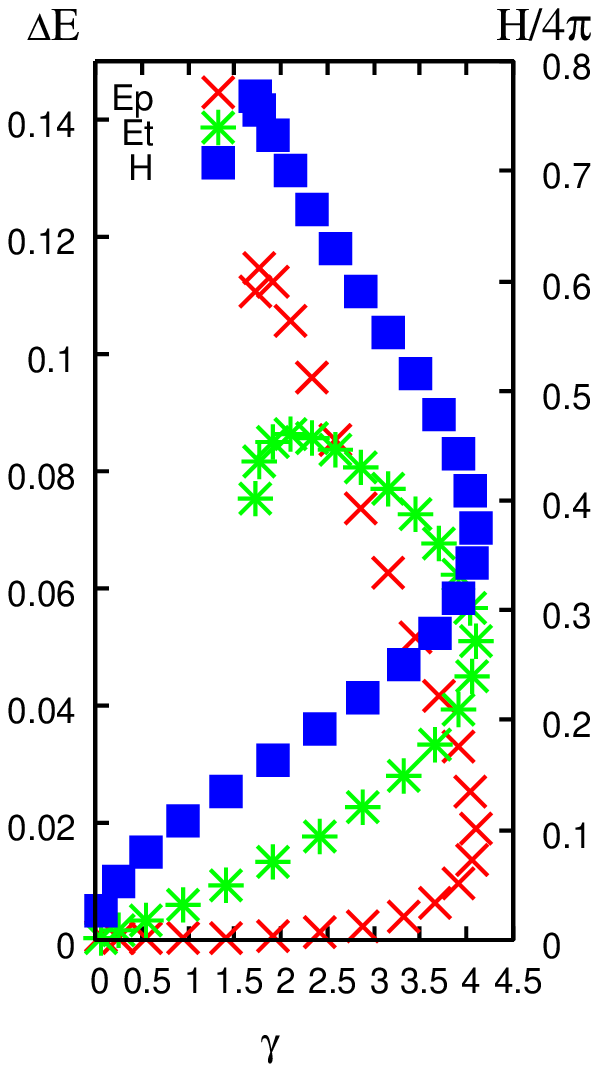}
\includegraphics[scale=0.45]{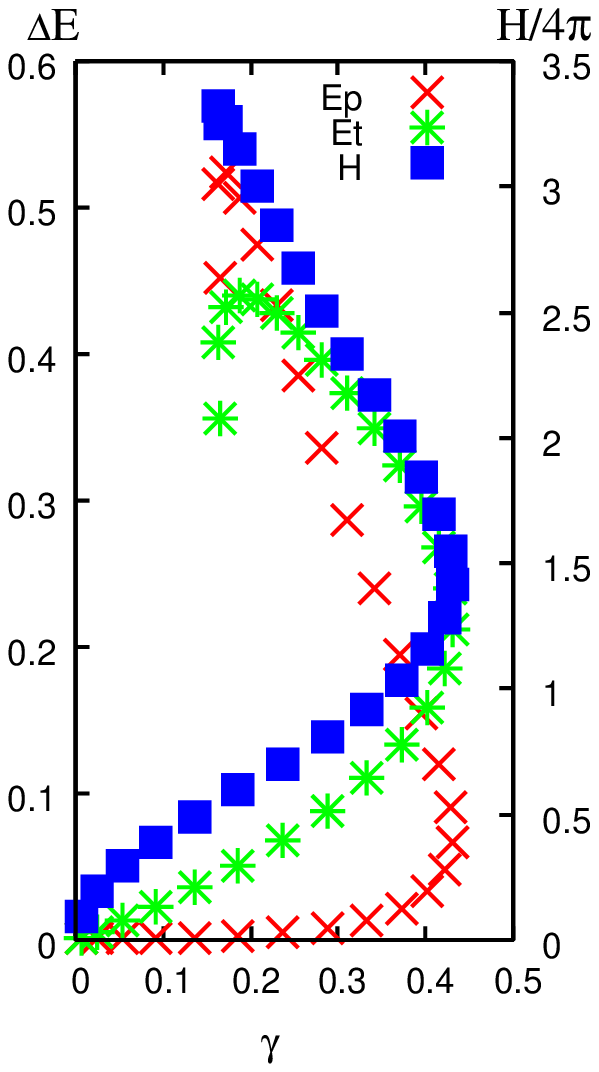} 
\includegraphics[scale=0.45]{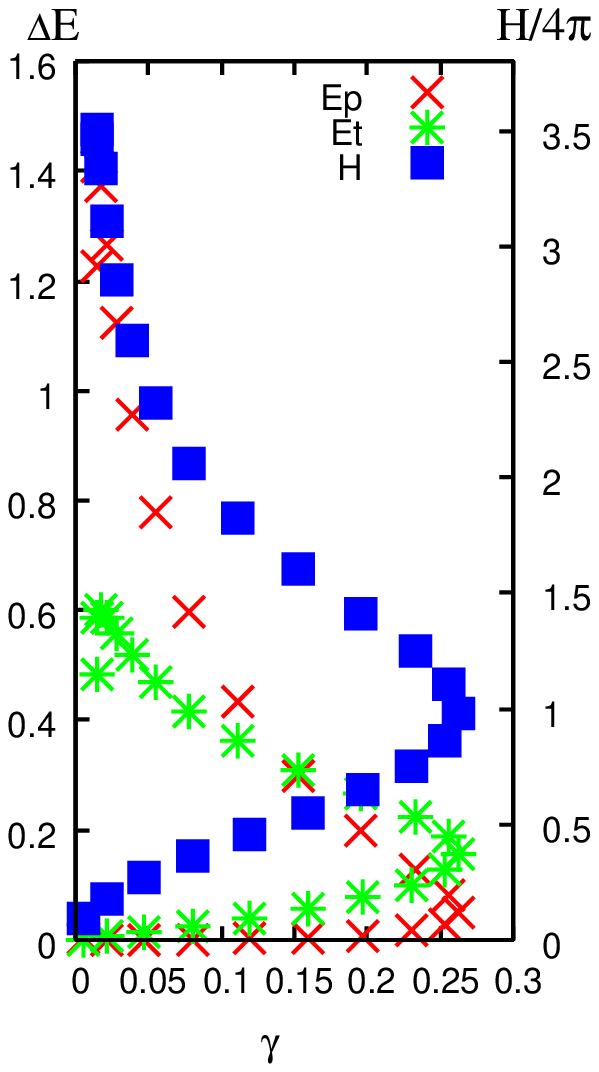} 
\caption{Increase in magnetic energy $\Delta E/(B_0^2 R^3)$ 
from the potential dipole field is shown in the left scale and
total relative helicity $H_{\rm R} /(4\pi B_0^2 R^4)$
is shown in the right scale.
The poloidal component of the energy is
denoted by crosses, the toroidal  component by asterisks,
and helicity by squares.
The horizontal axis denotes 
the dimensionless value $\gamma(B_0 R^2)^n $.
From left to right, the panels show the results for the 
F5, F7, G5 and G7 models in Table \ref{Table:I}.
}
\label{fig:1}
\end{figure}

 We first construct a magnetosphere model with a dipole in vacuum, 
and gradually increase the toroidal component for a given
power-law index $n$ and a relativistic factor $M/R$.
Calculations in flat spacetime \citep{2004ApJ...606.1210F}  
indicate that a free parameter $\gamma$ is not suitable for specifying 
the sequence of models, because there are multiple solutions 
for a given $\gamma$.
Instead, the azimuthal magnetic flux or 
helicity is used as the degree of twist, and the constant $\gamma$ 
is posteriorly determined.
We follow the same method, and calculate the energy and the
relative helicity along the sequence of solutions.
The characteristics of the models are given in Table \ref{Table:I}.
The increase in magnetic energy
and the relative helicity are shown for four models in Fig. \ref{fig:1}. 
The results in flat spacetime
agree with those by \citet{2004ApJ...606.1210F,2006ApJ...644..575Z}.
For a better understanding of the mechanism, the energy 
difference $\Delta E(=E_{\rm EM}-E_{0})$ is divided into 
$\Delta E =\Delta E_{t}+\Delta E_{p}$,
that of toroidal component and that of the poloidal component.
With increasing helicity from zero in the weak regime, the
energy $\Delta E_{t}$ increases monotonically, while
the energy $\Delta E_{p}$ is almost zero.
This phase, which corresponds to the lower branch of the sequence
in Fig. \ref{fig:1}, is characterized by increasing 
strength parameter $\gamma$. 
There is a maximum of $\gamma$, and after passing the turning point,
the energy $\Delta E_{p}$ drastically increases
in the upper branch in Fig. \ref{fig:1}.
This means a significant structure change from 
that in the dipole potential field. This fact will be confirmed later.
The numerical calculation becomes complicated 
for finding equilibrium solutions with larger helicity.
This terminates our numerical calculations.
The endpoint of the sequence is mostly consistent with that of 
\cite{2004ApJ...606.1210F,2006ApJ...644..575Z}.
It is likely that a static configuration no longer exists, leading to
a catastrophic transition associated with ejection of magnetic energy upon further
accumulation of helicity.

In the relativistic treatment, the increase of magnetic energy
is much larger. This does not merely come from 
the normalization of the surface field.
As discussed in the previous section, the surface field  
in the relativistic model
is a few times larger than in the flat spacetime model
with the same magnetic dipole moment $\mu$. 
Therefore, $\Delta E$ itself is naturally scaled up.
The ratio of $\Delta E$ to $E_{0}$,
which does not depend on the normaliztion,
is also significantly increased.
The maximum of each model is given in Table \ref{Table:I}. 
The curved spacetime allows higher energy and helicity 
to be stored in the magnetosphere.
Nonlinearity with respect to the index $n$ also enhances
the maximum energy. For example, the maximum in the model
with $n=7$ is larger than for $n=5$.
The relativistic correction, however, is much larger.

\begin{figure}
\centering
\includegraphics[scale=0.65]{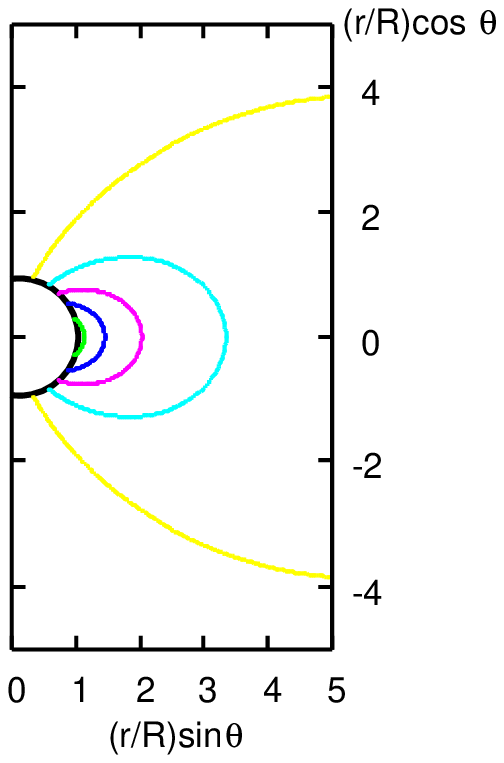}
\includegraphics[scale=0.65]{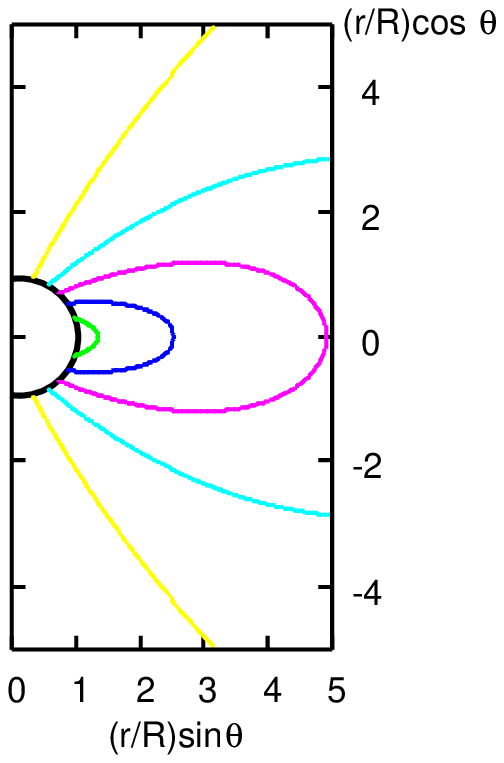}
\includegraphics[scale=0.65]{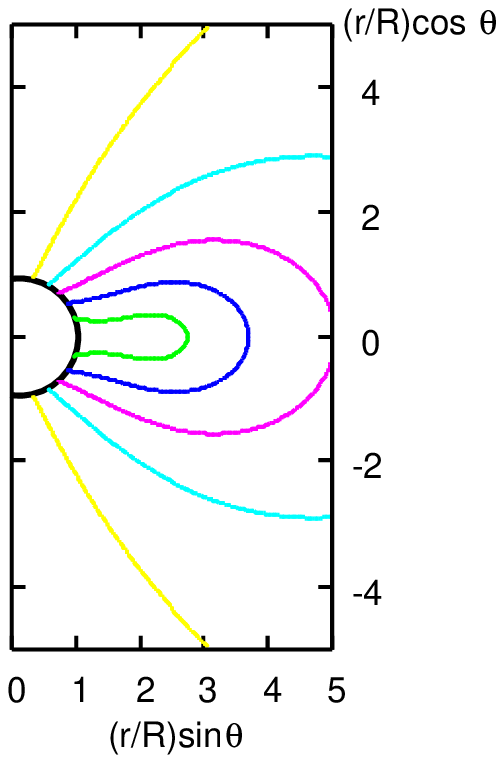} 
\caption{Contour of magnetic function $G$
in the $r$-$\theta$ plane.
Left panel shows the potential dipole field,
middle shows the highly twisted state with $n=5$,
and right shows the highly twisted state with $n=7$.
Contour lines represent the level of $G$
for $0.1 \times (2k+1) \times (B_{0} R^{2}), k=0, 1,2,\cdots$.
}
\label{fig:2}
\end{figure}

%
Figure \ref{fig:2} shows the magnetic function $G$ obtained 
in flat spacetime by contours in the $r$-$\theta$ plane for the inner region $(r/R \le 5)$.
The magnetic field structure with large helicity is
significantly changed, as shown in the second and third panels.
By increasing the helicity,
the magnetic field lines become stretched outwardly
and slightly squashed into the equator.
\citet{2004ApJ...606.1210F,2006ApJ...644..575Z} 
discovered an interesting property of the highly twisted state:
A local maximum of $G$ exists 
just outside the star at the equator.
The peak, which becomes more evident for larger 
$n$,  is located inside the stretched poloidal field.
This is numerically confirmed in our model, but the
peak is too small to appear in 
the coarse interval of the contour levels.

\begin{figure}
\centering
\includegraphics[scale=0.65]{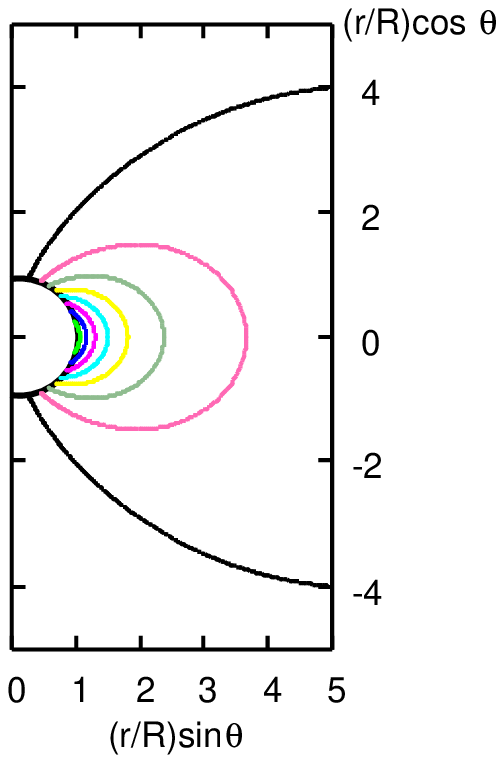}
\includegraphics[scale=0.65]{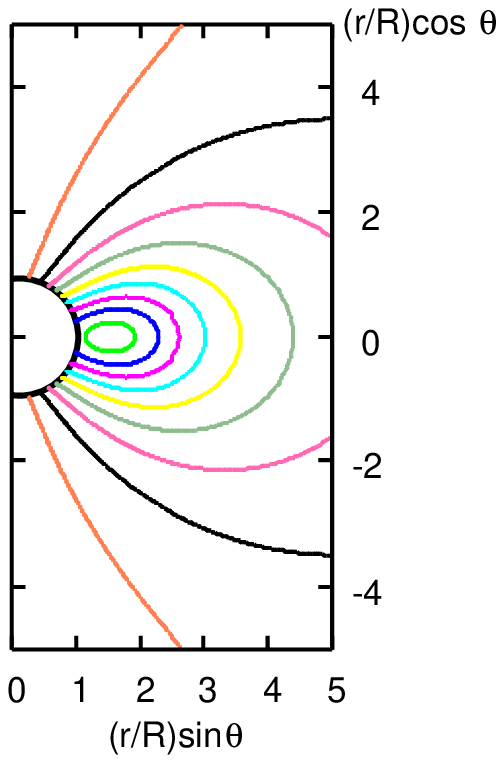}
\includegraphics[scale=0.65]{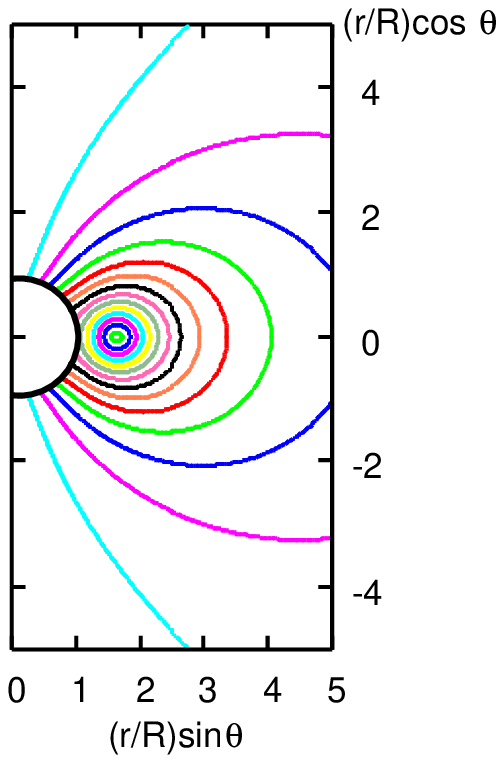}
\caption{
The same as Fig. \ref{fig:2}, but for the model with a
relativistic factor $M/R =0.25$.
}
\label{fig:3}
\end{figure}

Results for the relativistic treatment are shown in Fig. \ref{fig:3}. 
As discussed with respect to eq. (\ref{eqnpuredip}), the magnetic field strength
near the surface is increased
by a few factors for the same magnetic dipole moment $\mu$.
The structure of the potential magnetic dipole shown in the left panel 
is the same as obtained
in flat spacetime except for the magnitude.
However, there are evident loops of the field lines
in models with large helicity, as shown the second and third panels in Fig. \ref{fig:3}.
This structure represents a magnetic flux rope with field line
projections on the $r$-$\theta$ plane 
in the form of a closed curve of a constant $G$.
The center of the loops ($r/R \approx 1.5$, $\theta =\pi/2$)
corresponds to a global maximum of $G$,
whereas the center of potential dipole field $G_{\rm d}$  
is located at the stellar surface, $r=R$ and $\theta =\pi/2$.
The loop structure is clearly evident in the model with $n=7$
due to the stronger nonlinearity of $n$. 
It should be noted that the loop structure is 
formed in the upper branch of the sequence in Fig. \ref{fig:1}.
It has higher energy and helicity for a fixed $\gamma$.
It is therefore very important to know whether this kind of higher energy state
appears evolutionarily. 
A helically twisted flux rope is thought to be a phase that precedes
magnetar flares \citep{2011ApJ...738...75Y,2014ApJ...796....3H}.
The structure is modeled by a small ring current 
at $(r,\theta)=(r_{0}, \pi /2 )$.
The results show that for a stable equilibrium solution 
the radial distance $r_{0}$ is limited to a few times the stellar radius.
Our present result is an explicit example of a fixed rope.

\begin{figure}
\centering
\includegraphics[scale=1.0]{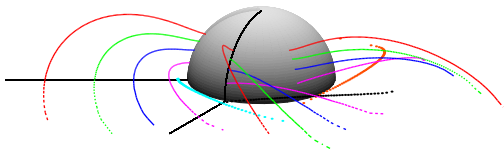} 
\caption{Magnetic field lines in the upper half-plane
($0\le \theta\le \pi/2$).
Four lines starting from the stellar surface at azimuthal angles
$\phi =0, \pi/3, 2\pi/3$ are displayed.
Two angles $\phi =0$ and $\pi/3$
are indicated by lines.
Field lines on the torus are also displayed.
They start from $\phi =0, \pi/3$, and $2\pi/3$ on the
equator return to it.}
\label{fig:4} 
\end{figure}

%
Figure \ref{fig:4} shows a 3-dimensional structure 
for the highly twisted relativistic model with $n=5$.
Twisted lines anchored to the central star are shown
in only  the hemisphere ($0 \le \theta \le \pi/2$).
The figure also shows that the typical twist angle $\Delta \phi $
is 1 radian, where $\Delta \phi $ denotes
the angle from the stellar surface to the equator  
in eq. (\ref{eqn:wistang}).
It can be easily understood that the maximum of 
$\Delta \phi $ for a magnetosphere model is given by a line 
starting from the surface with an intermediate latitude
$\theta_{*} $ ($0 < \theta_{*} < \pi/2$).
However, the exact value $\theta_{*} $ depends on the structure, and 
numerical work is necessary to find it
by calculating $\Delta \phi $ for all possible lines. 
We search the maximum among calculated sequences of
four models.
Their values are tabulated in Table \ref{Table:I}, and 
are all approximately $\Delta \phi_{\rm max} \approx 1$ radian.
It should be noted that the maximum in the model sequence
is not necessarily attained in the state 
with the largest helicity.
The maximum in the G7 model is smaller, 
although a large amount of helicity can be stored. 
This fact is related to the loop formation.
When this kind of detached structure forms,
the lines starting from the star and returning to it
do not extend radially. The twist 
in the azimuthal direction becomes small. 
The maximum twist $\Delta \phi_{\rm max} =$0.8-1.5
in our models is interesting by comparison
with a different model:
$\Delta \phi_{\rm max} =1.2$-1.5 in
the model of \citet{2016MNRAS.462.1894A},
and $\Delta \phi_{\rm max} \le \pi /2$ in a self-similar model
\citep{2002ApJ...574..332T}
\footnote{
Our twisted angle $\Delta \phi $ is
defined between a point in north-hemisphere to equator.
The twisted angle $\Delta \phi_{\rm NS}$ in 
\citep{2002ApJ...574..332T} 
is $\Delta \phi_{\rm NS}=2\Delta \phi$.
}.
Finally, it should be noted that, in a sense,
the quantity $\Delta \phi $ represents
the degree of twisted structure, but
is not necessarily a good indicator 
because it is not a monotonic function along a sequence.

\section{Conclusion}
  In this paper, we have studied how a force-free magnetosphere is 
changed by the accumulation of helicity. The model of current is 
specified by a simple power law  by which
the equilibria in flat spacetime are extensively examined 
in the context of solar flares.  We extend the study to 
the general relativistic regime for application to magnetar flares.
A family of axially symmetric static solutions is calculated.
Total magnetic energy increases with increasing helicity.
Numerical calculations suggest that there is a maximum of
the helicity deposited in the magnetosphere. 
Beyond this, we could not find any static solutions.
The existence of a critical state, which is characterized
by magnetic helicity or twist angle in 
\citet{2002ApJ...574..332T,2006ApJ...644..575Z,2016MNRAS.462.1894A},
is also found in the 
different models and by different numerical approaches.
This fact is very interesting for the eruption of magnetic flux rope.
When the total helicity injected into the magnetosphere
exceeds a certain capacity, a catastrophic event such as a flare
may occur.
This kind of sudden event may be calculated by 
resistive dynamical 
simulation \citep{2012ApJ...746...60L,2013ApJ...774...92P,2014PTEP.2014b3E01K}.
It is also important to determine when
an instability sets in along our equilibrium sequence:
at the endpoint of our calculation 
or at a certain state before it.
  We found that general relativistic effects are significant.
A larger amount of energy is capable of being stored in the relativistic 
magnetosphere. In extreme cases, the increase in magnetic energy 
in the presence of current flow exceeds that of a current-free dipole.
This marks a contrast with the models in flat spacetime, 
in which the magnetic energy increment
is less than that of vacuum dipole.
When the magnetic field structure is highly twisted, 
a loop of field lines in 2-dimensional meridian plane, 
which corresponds to an axially symmetric torus in 
3-dimensional space, is formed in the vicinity of the surface.
This kind of remarkable structure is also realized in flat spacetime,
but the condition to the power-law index $n$ is severe.
Namely, evident structure is realized only in the model with stronger 
nonlinearity, $ n \ge 7$ 
\citep{2004ApJ...606.1210F,2006ApJ...644..575Z}.
The loops are also found in the exterior model of a neutron star 
obtained by solving the relativistic MHD equation
with a low-density plasma \citep{2015MNRAS.447.2821P}.
A direct comparison is difficult due to the differences 
in both models and 
numerical methods, but general relativity is likely to play 
an important role in the formation.
  Observationally, total magnetic energy in the magnetar magnetosphere
is approximately $\sim 10^{48} (B_0/10^{15} {\rm G})^2 (R/10 {\rm km})^3$.
The flare energy is about 1\% of this energy.
This means that the flare is associated with a small change 
between the equilibrium models. 
From the point of the energy scale,
there are two possibilities:
a small transition between twisted dipolar equilibria or
a transition of a higher multipolar field 
with weaker strength.
Long-term evolution of the internal magnetic field
by Hall drift demonstrates that this kind of higher multipolar field
emerges at the surface \citep[e.g.,][]
{2004MNRAS.347.1273H,2012MNRAS.421.2722K,
2013MNRAS.434..123V,2014MNRAS.438.1618G,2015PhRvL.114s1101W}, 
although the results depend on 
the initial conditions.
A correct functional form $S=S(G)$ at the surface,
which is important for constructing force-free magnetosphere, 
will be determined by connecting internal evolution.


\section*{Acknowledgements}
This work was supported in part by a Grant-in-Aid for Scientific Research
(No. 25103414) from the Japanese Ministry of Education, Culture, Sports,
Science and Technology.

 \bibliographystyle{mnras}
 \bibliography{kojima17Feb} 

\end{document}